\documentclass[12pt]{article}
\usepackage{graphicx}
\usepackage{amsmath}
\usepackage{psfrag}
\usepackage{amssymb}
\usepackage{fullpage}
\usepackage{subfigure}

\begin{document}




\title{\bf Probability distribution of the resistance \\of a random network}
\setcounter{footnote}{2}

\author{Thomas Callaghan$^*$ and
  Joseph B. Keller$^*$ $^{\dagger}$ \bigskip \\
  $^*$Institute for Computational and Mathematical Engineering\\
  	Stanford University, Stanford, CA 94305-4042 \bigskip \\
  $^\dagger$Department of Mathematics \\
	Stanford University, Stanford, CA 94305-2125}
\maketitle

\begin{abstract}
The probability density of the resistance of a two dimensional rectangular network between two conducting plates is calculated.  The nodes form an $M$ by  $N$ lattice, and each edge has a random resistance.  The Monte Carlo method is used.
\end{abstract}

\section{Introduction}
\setcounter{equation}{0}
A random electrical network is a network of resistors, with the resistance of each resistor being a random variable.  Then the overall resistance of the network between any two nodes is also a random variable.  Its probability distribution is determined by the probability distributions of the individual resistors, and by the structure of the network. 

We shall calculate the probability distribution  of the resistance of a two dimensional rectangular network between two conducting plates.  The nodes of the network form an $M$ by $N$ lattice, and each edge is a resistor.  The resistances of the resistors are independent identically distributed ({\em i.i.d.\/}) random variables, taking the value $r\leq 1$ with probability $p$, and the value $1$ with probability $1-p$.  See Figure 1.

In the special case $M=1$, the network is just a line of $N$ resistors in series.  Its resistance is $R=\sum\limits^N_{j=1} \,r_j$, where $r_j$ is the $j$-th resistance.  Thus the probability distribution of $R$ is just that of the sum of $N$ {\em i.i.d.\/} random variables, which is a binomial distribution with parameters $N$ and $p$.  In the present case the possible values of $R$ and their probabilities are
\begin{equation}
R_k = kr + \left( N-k\right) ,\qquad P\left(R_k\right) = \binom{N}{p} \, p^k \left( 1-p\right)^{N-k}, \\\quad k=1,\cdots,N.
\label{eq:1.1}
\end{equation}
\section{Formulation for $M>1$}
\setcounter{equation}{0}
For $M>1$, we must determine $R$ by first solving Kirchoff's equations for the currents and potentials.  Let $i,j$ denote the $(j+1)$-st node on the horizontal line $i$, with $j=0$ on the plate at the left and $j=N$ at the plate on the right.  Let $\phi_{ij}$ be the potential at node $i,j$.  Finally let $r_{ij, (i+1)j}$ be the resistance of the resistor from $i,j$ to $i+1, j$, etc., and let $I_{ij,(i+1)j}$ be the current in that resistor.  Then Kirchoff's equations can be written as
\begin{equation}
\begin{split}
\phi_{i,j+1}-\phi_{i,j}&=I_{ij,i(j+1)}r_{ij,i(j+1)} \\
\phi_{i+1,j}-\phi_{i,j}&=I_{ij,(i+1)j}r_{ij,(i+1)j} \\
I_{(i-1)j,ij}+I_{i(j-1),ij} &= I_{ij,(i+1)j}+I_{ij,i(j+1)}
\end{split}
\label{eq:2.1}
\end{equation}

We must solve (\ref{eq:2.1}) for the currents and potentials, subject to the boundary conditions
\begin{equation}
\phi_{i0} =0, \qquad \phi_{iN}=1.
\label{eq:2.2}
\end{equation}
Then the total current from the plate at the left to that at the right is:
\begin{equation}
I=\sum\limits^M_{i=1} I_{i0, i1}.
\label{eq:2.3}
\end{equation}
The resistance between the plates is $R=I^{-1}$, which is the random variable, the distribution of which we seek.  We normalize it by setting $\bar{R}=MR/N$.  Then $r\leq \bar{R}\leq 1$.

To find the probability distribution  $P(\bar{R})$ we use the Monte Carlo method.  This consists in generating a sample network by selecting each resistance to be $r$ with probability $p$, or $1$ with probability $1-p$, solving (\ref{eq:2.1}) and (\ref{eq:2.2}), and calculating $I$ from (\ref{eq:2.3}).  Then we obtain the sample value:  $\bar{R} =MI^{-1}/N$.  By repeating this process many times, we generate a large set of sample values. We sort them into bins, which are subdivisions  of the interval $(r, 1)$ of the $\bar{R}$ axis.  The fraction of samples in each bin, divided by the bin length, is the empirical probability of $\bar{R}$ at the midpoint of the bin.  The mean of the sample values is the empirical expected value $E(\bar{R})$.

The network has four parameters $r$, $p$, $N$, and $M$.  In our calculations we set $M=N$ so $\bar{R} =R$.  Only the three parameters $r$, $p$ and $N$ remain.

\section{Results}
\setcounter{equation}{0}
We calculated $P(R)$ and $E(R)$ for $N=2, 10, 20, 30$; $r=.25, .50, .75$ and $p=.25, .50, .75$.  In each case we used 2000 samples, and the calculated $P(R)$ was close to a normal distribution, being closer the larger the value of $N$.  Furthermore the expected value was close to the power law, $E(R)\approx r^p$, again becoming closer as $N$ increased.  Some of the calculated results are shown in Figures 2--5.

In Figure 2, the calculated  $P(R)$ is shown together with a closely fitting normal distribution for $r=1/2$, $p=1/2$ nd $N=10, 20, 30$.  In Figure 3a, the calculated $E(R)$ is shown as a function of $p$ for $N=20$ and three values of $r$.  In Figure 3b, $E(R)$ is shown as a function of $r$ for $N=20$ and three values of $p$.  In each case the function $r^p$ is also shown.

In Figure 4a, the calculated $E(R)$ is shown as a function of $p$ for $N=2$ and $r=.5$ together with $r^p$, and the exact result for $E(R)$.  The latter was obtained by solving (\ref{eq:2.1}) and (\ref{eq:2.2}) for $R$ using {\em Mathematica\/}, and then obtaining $E(R)$ from the solution.  The result is given in the appendix.  Figure 4b is similar with $p=.5$ and $r$ varying.

\section*{Acknowledgements}

We thank Chris Maes for some useful discussions and help with Mathematica.

\newpage
\section*{Appendix}
\begin{multline*}E[R]={\left( 1 - p \right) }^5 + {\left( 1 - p \right) }^4\,p + p^5\,r +
  \frac{4\,{\left( 1 - p \right) }^4\,p\,\left( 3 + 5\,r \right) }{5 + 3\,r} +
  \frac{{\left( 1 - p \right) }^2\,p^3\,\left( 3\,r + 4\,r^2 + r^3 \right) }
   {1 + 4\,r + 3\,r^2} \\ + \frac{\left( 1 - p \right) \,p^4\,
     \left( 5\,r^2 + 3\,r^3 \right) }{3\,r + 5\,r^2} +
  \frac{{\left( 1 - p \right) }^3\,p^2\,
     \left( 1 + 4\,r + r^2 + r\,\left( 1 + r \right)  \right) }{2 +
5\,r + r^2} \\ +
  \frac{{\left( 1 - p \right) }^3\,p^2\,
     \left( 1 + 4\,r + r^2 + r\,\left( 1 + r \right)  \right) }{2 + 4\,r +
     r\,\left( 1 + r \right) }+ \frac{{\left( 1 - p \right) }^3\,p^2\,
     \left( 1 + 2\,r + r^2 + 2\,r\,\left( 1 + r \right)  \right) }{3 + 3\,r +
     r\,\left( 1 + r \right) } \\ + \frac{{\left( 1 - p \right) }^3\,p^2\,
     \left( 1 + 4\,r + 3\,r^2 \right) }{3 + 2\,r + r\,\left( 2 + r \right) } +
  \frac{2\,{\left( 1 - p \right) }^3\,p^2\,
     \left( 1 + 2\,r + 2\,r^2 + r\,\left( 2 + r \right)  \right) }{2 + 6\,r} \\ +
  \frac{{\left( 1 - p \right) }^3\,p^2\,
     \left( 1 + 2\,r + r\,\left( 1 + r \right)  + r\,\left( 2 + r
\right)  \right) }
     {2 + 4\,r + r\,\left( 1 + r \right) } +
  \frac{{\left( 1 - p \right) }^3\,p^2\,
     \left( 1 + 2\,r + r\,\left( 1 + r \right)  + r\,\left( 2 + r
\right)  \right) }
     {2 + 3\,r + r\,\left( 2 + r \right) } \\ +
  \frac{{\left( 1 - p \right) }^3\,p^2\,
     \left( 3\,r + r\,\left( 1 + r \right)  + r\,\left( 2 + r \right) 
\right) }{3 +
     3\,r + r\,\left( 1 + r \right) } +
  \frac{{\left( 1 - p \right) }^3\,p^2\,
     \left( 3\,r + r\,\left( 1 + r \right)  + r\,\left( 2 + r \right) 
\right) }{3 +
     2\,r + r\,\left( 2 + r \right) } \\ +
  \frac{{\left( 1 - p \right) }^2\,p^3\,
     \left( r + 2\,r^2 + r\,\left( 1 + r \right)  + r^2\,\left( 2 + r
\right)  \right) }
     {1 + 2\,r + r\,\left( 1 + r \right)  + r\,\left( 2 + r \right) } +
  \frac{{\left( 1 - p \right) }^2\,p^3\,
     \left( r + 2\,r^2 + r\,\left( 1 + r \right)  + r^2\,\left( 2 + r
\right)  \right) }
     {1 + 4\,r + r\,\left( 1 + 2\,r \right) } \\ +
  \frac{\left( 1 - p \right) \,p^4\,\left( 5\,r^2 + 3\,r^3 \right) }
   {2\,r + 3\,r^2 + r\,\left( 1 + 2\,r \right) } +
  \frac{{\left( 1 - p \right) }^2\,p^3\,
     \left( r + 2\,r\,\left( 1 + r \right)  + r^2\,\left( 2 + r
\right)  \right) }{1 +
     2\,r + r\,\left( 1 + r \right)  + r\,\left( 1 + 2\,r \right) } \\ +
  \frac{{\left( 1 - p \right) }^2\,p^3\,
     \left( 3\,r^2 + r\,\left( 1 + r \right)  + r\,\left( 1 + 2\,r
\right)  \right) }
     {1 + 4\,r + 3\,r^2} + \frac{{\left( 1 - p \right) }^2\,p^3\,
     \left( 3\,r^2 + r\,\left( 1 + r \right)  + r\,\left( 1 + 2\,r
\right)  \right) }
     {1 + 2\,r + r\,\left( 1 + r \right)  + r\,\left( 1 + 2\,r \right) } \\ +
  \frac{{\left( 1 - p \right) }^2\,p^3\,
     \left( 2\,r + r^2 + r^2\,\left( 1 + r \right)  + r\,\left( 1 +
2\,r \right)  \right)
       }{4\,r + 2\,r\,\left( 1 + r \right) } +
  \frac{{\left( 1 - p \right) }^2\,p^3\,
     \left( 2\,r + r^2 + r^2\,\left( 1 + r \right)  + r\,\left( 1 +
2\,r \right)  \right)
       }{4\,r + r^2 + r\,\left( 2 + r \right) } \\ +
  \frac{{\left( 1 - p \right) }^2\,p^3\,
     \left( r^2 + r\,\left( 1 + r \right)  + r^2\,\left( 1 + r \right)  +
       r\,\left( 1 + 2\,r \right)  \right) }{1 + 3\,r + 2\,r\,\left( 1
+ r \right) } \\ +
  \frac{{\left( 1 - p \right) }^2\,p^3\,
     \left( r^2 + r\,\left( 1 + r \right)  + r^2\,\left( 1 + r \right)  +
       r\,\left( 1 + 2\,r \right)  \right) }{1 + 2\,r + r\,\left( 1 +
r \right)  +
     r\,\left( 2 + r \right) } + \frac{\left( 1 - p \right) \,p^4\,
     \left( 3\,r^2 + r^2\,\left( 1 + r \right)  + r^2\,\left( 1 + 2\,r
\right)  \right) }
     {r + 3\,r^2 + 2\,r\,\left( 1 + r \right) } \\ + 
  \frac{\left( 1 - p \right) \,p^4\,
     \left( 3\,r^2 + r^2\,\left( 1 + r \right)  + r^2\,\left( 1 + 2\,r
\right)  \right) }
     {2\,r + 3\,r^2 + r\,\left( 1 + 2\,r \right) } +
  \frac{\left( 1 - p \right) \,p^4\,
     \left( r^2 + 2\,r^2\,\left( 1 + r \right)  + r^2\,\left( 1 + 2\,r
\right)  \right) }
     {r + 2\,r\,\left( 1 + r \right)  + r\,\left( 1 + 2\,r \right) }~.\end{multline*}

\newpage

\begin{center}
{\bf Figure Captions }
\end{center}

\begin{enumerate}
\item A two dimensional network with $N=10$ rows and $M=10$ columns of resistors.  At the left side is a plate with potential $\phi_{i,0}=0$ and at the right side a plate with $\phi_{i,10}=1$.
\item Simulated probability density $P(R)$ using 2000 samples, together with a closely fitted normal density for $r=.5$, $p=.5$ and a) $N=10$, b) $N=20$, c) $N=30$.
\item Simulated mean $E(R)$ (crosses) using 2000 samples, together with the curve $r^p$, for $N=20$. a)~$r=.25$, $.5$, $.75$ with $p$ varying.  b) $p=.25$, $.5$, $.75$ with $r$ varying.
\item Simulated mean $E(R)$ (dotted line), $r^p$ (thin line) and exact $E(R)$ (heavy line) for $N=2$.  a) $r=.5$ with $p$ varying.  b) $p=.5$ with $r$ varying.
\item Simulated probability density $P(R)$ as a function of $R$ and $N$ with $r=.5$ and $p=.5$.
\end{enumerate}
\newpage
\begin{figure}
  {\psfragscanon\footnotesize
    \centerline{ \hspace{-0.15cm}
      \includegraphics[width=0.66\textwidth]{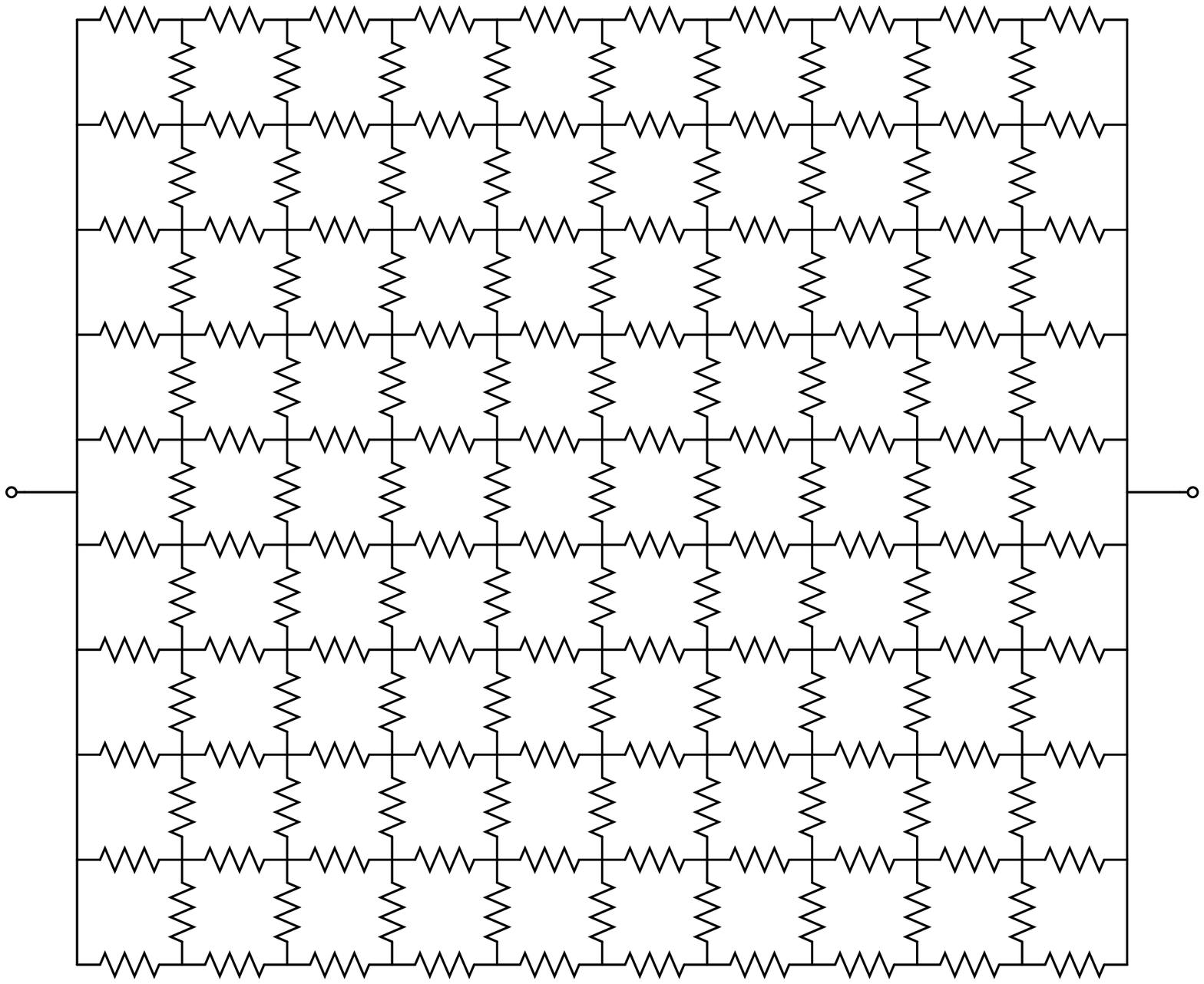}
    }
  }
  \caption{}
  \label{fig:network}
\end{figure}

\clearpage

\newpage
\vspace*{1in}
\begin{figure}[!htb]
  {\psfragscanon\footnotesize
    \psfrag{R}[t][b]{\vspace*{0.5cm}$R$}
    \psfrag{Empirical Probability}[][][0.7]{\footnotesize{Probability Density}}
    \psfrag{Simulation vs Gaussian, N=10}{\hspace*{-0.4in}Simulation vs Gaussian, N=10}
    \psfrag{Simulation vs Gaussian, N=20}{\hspace*{-0.4in}Simulation vs Gaussian, N=20}
    \psfrag{Simulation vs Gaussian, N=30}{\hspace*{-0.4in}Simulation vs Gaussian, N=30}
    \psfrag{frequency}{Frequency}
    \centerline{ (a)\hspace{-0.15cm}
      \includegraphics[width=0.35\textwidth]{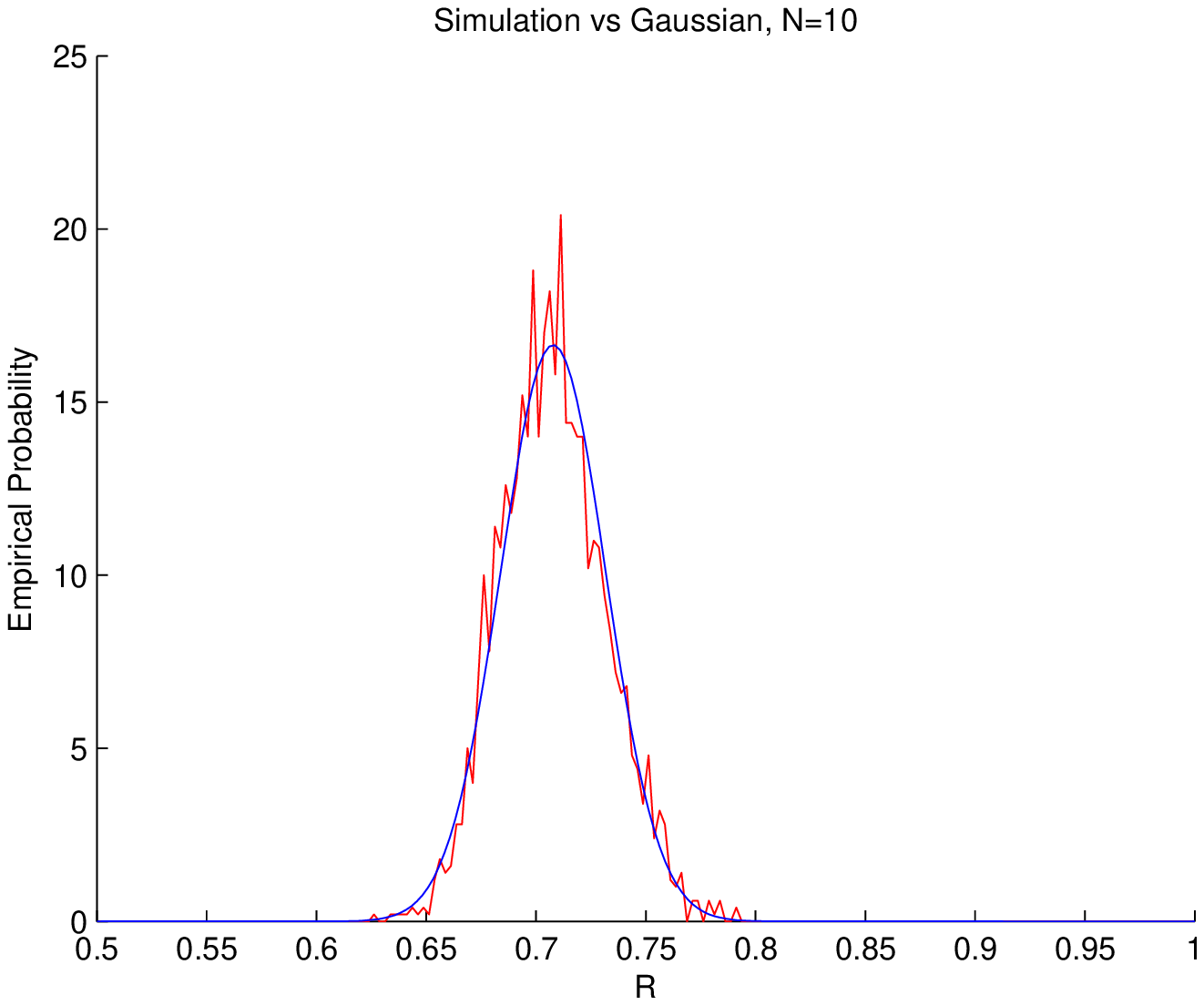}
      \hspace{0.2cm} (b)\hspace{-0.15cm}
      \includegraphics[width=0.35\textwidth]{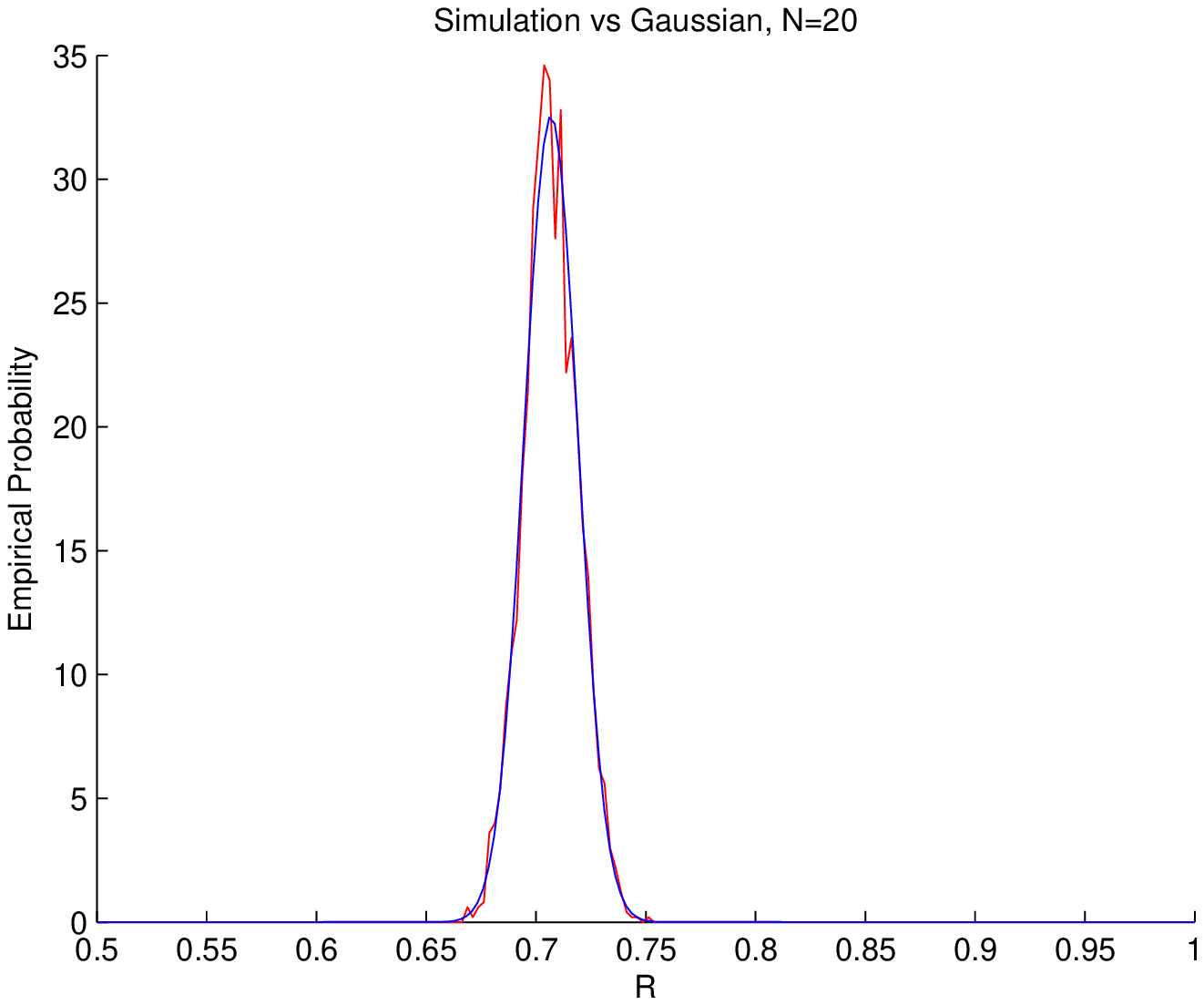}
      \hspace{0.2cm} (c)\hspace{-0.15cm}
      \includegraphics[width=0.35\textwidth]{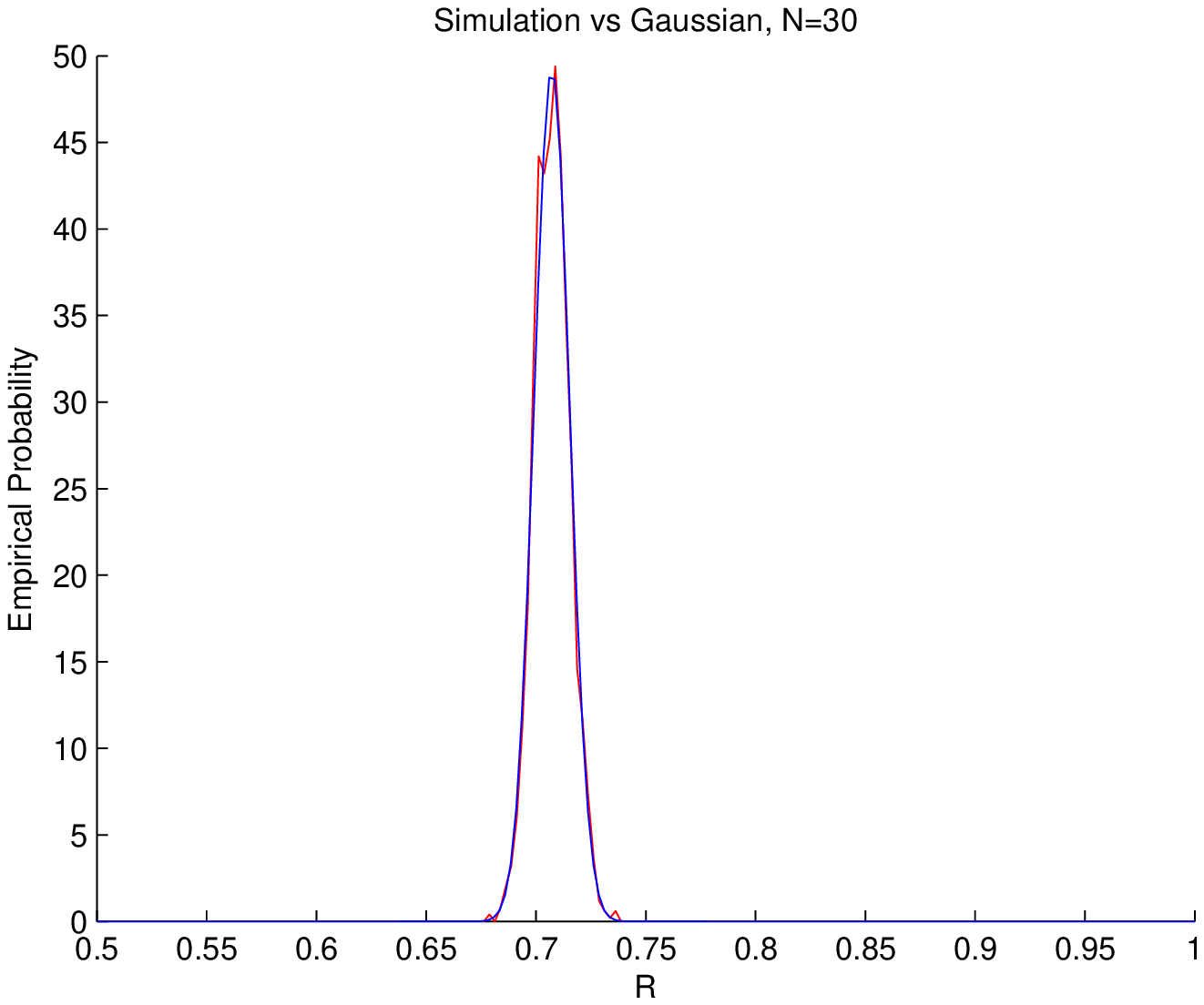}
    }
  }
  \caption{}
  \label{fig:gaussian}
\end{figure}

\begin{figure}[H]
  {\psfragscanon\footnotesize
    \psfrag{xlabel2}{$r$}
    \psfrag{xlabel}{$p$}
    \psfrag{E[R] for varied p, r = 0.25, 0.5, 0.75}{\hspace*{-0.4in}$E[R]$ for varied $p$, $r=.25, .5, .75$}
    \psfrag{E[R] for varied r, p = 0.25, 0.5, 0.75}{\hspace*{-0.4in}$E[R]$ for varied $r$, $p=.25, .5, .75$}
    \psfrag{ylabel}{$E[R]$}
    \centerline{ (a)\hspace{-0.15cm}
      \includegraphics[width=0.54\textwidth]{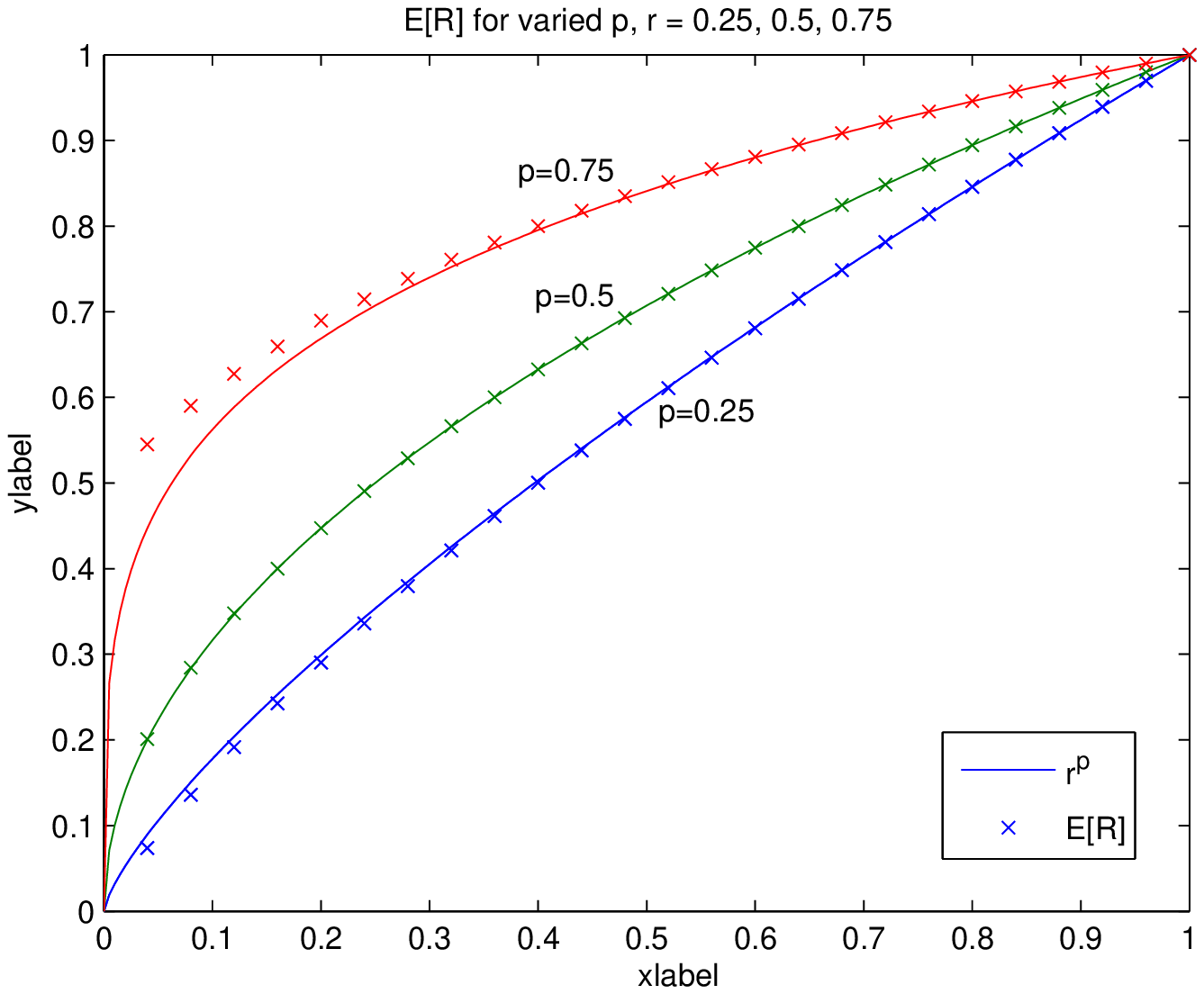}
      \hspace{0.2cm} (b)\hspace{-0.15cm}
      \includegraphics[width=0.54\textwidth]{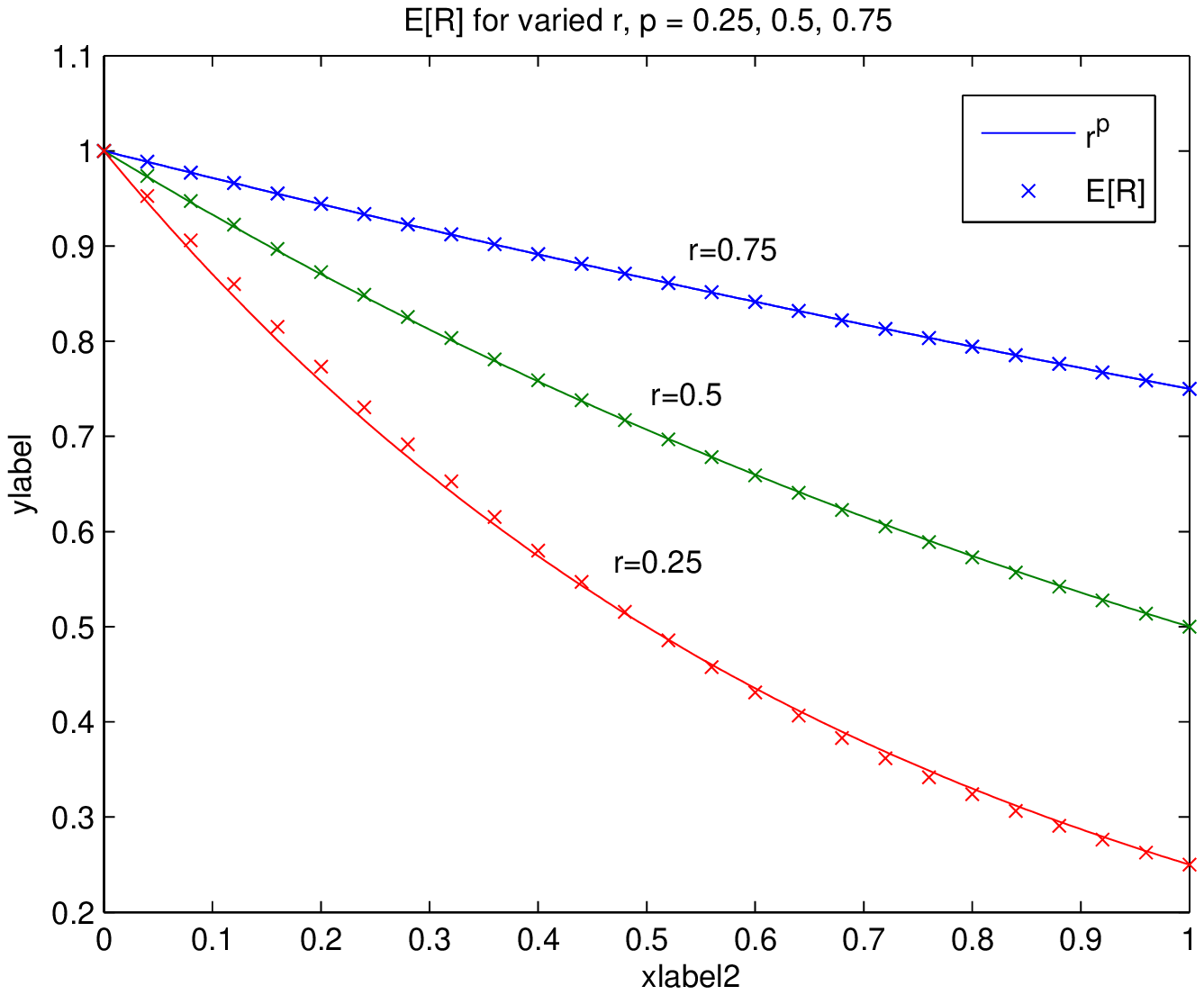}
    }
  }
  \caption{}
  \label{fig:varied}
\end{figure}

\begin{figure}[H]\centering
  {\psfragscanon\footnotesize
    \centerline{ (a)\hspace{-0.15cm}
      \includegraphics[width=0.54\textwidth]{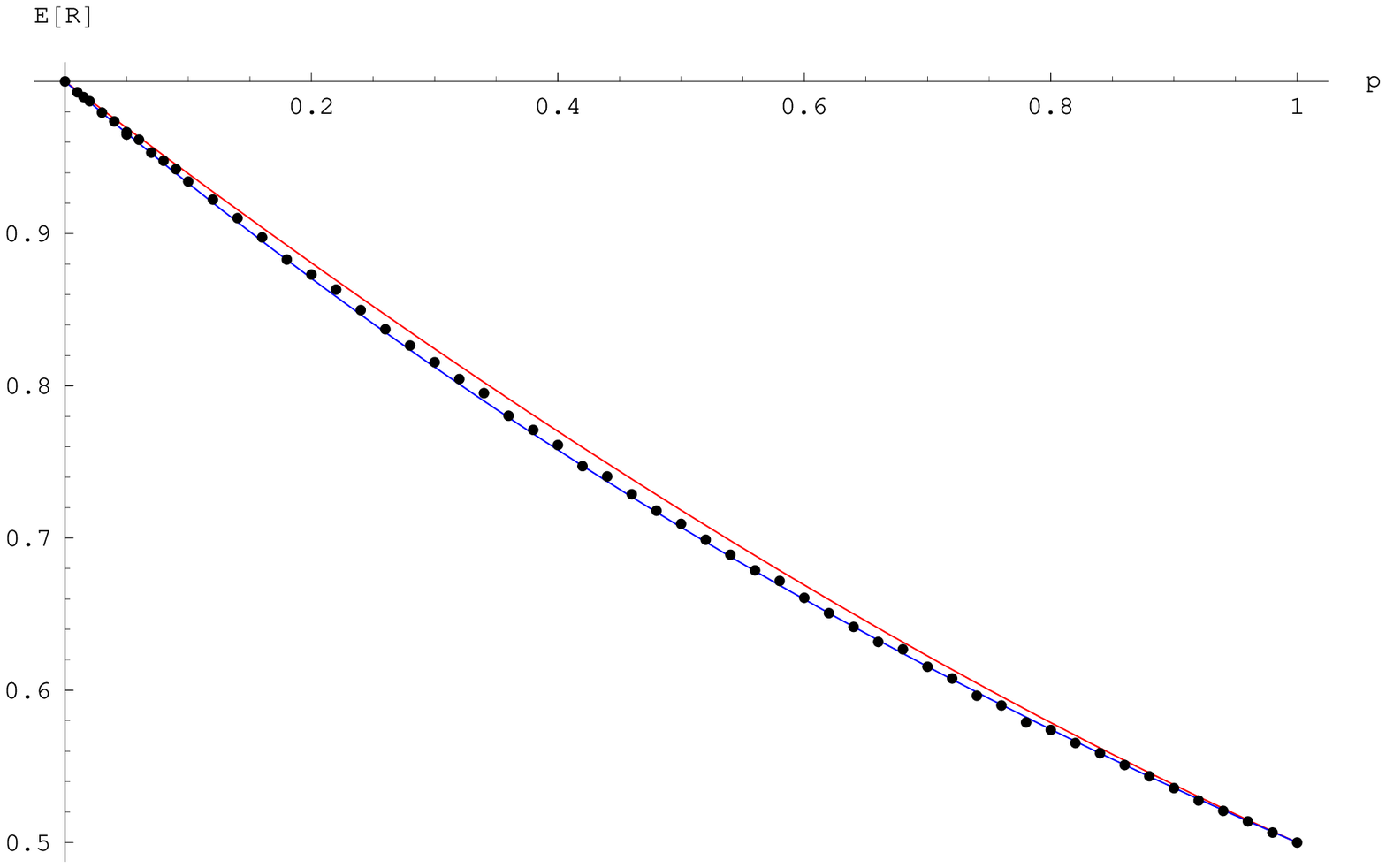}
      \hspace{0.2cm} (b)\hspace{-0.15cm}
      \includegraphics[width=0.54\textwidth]{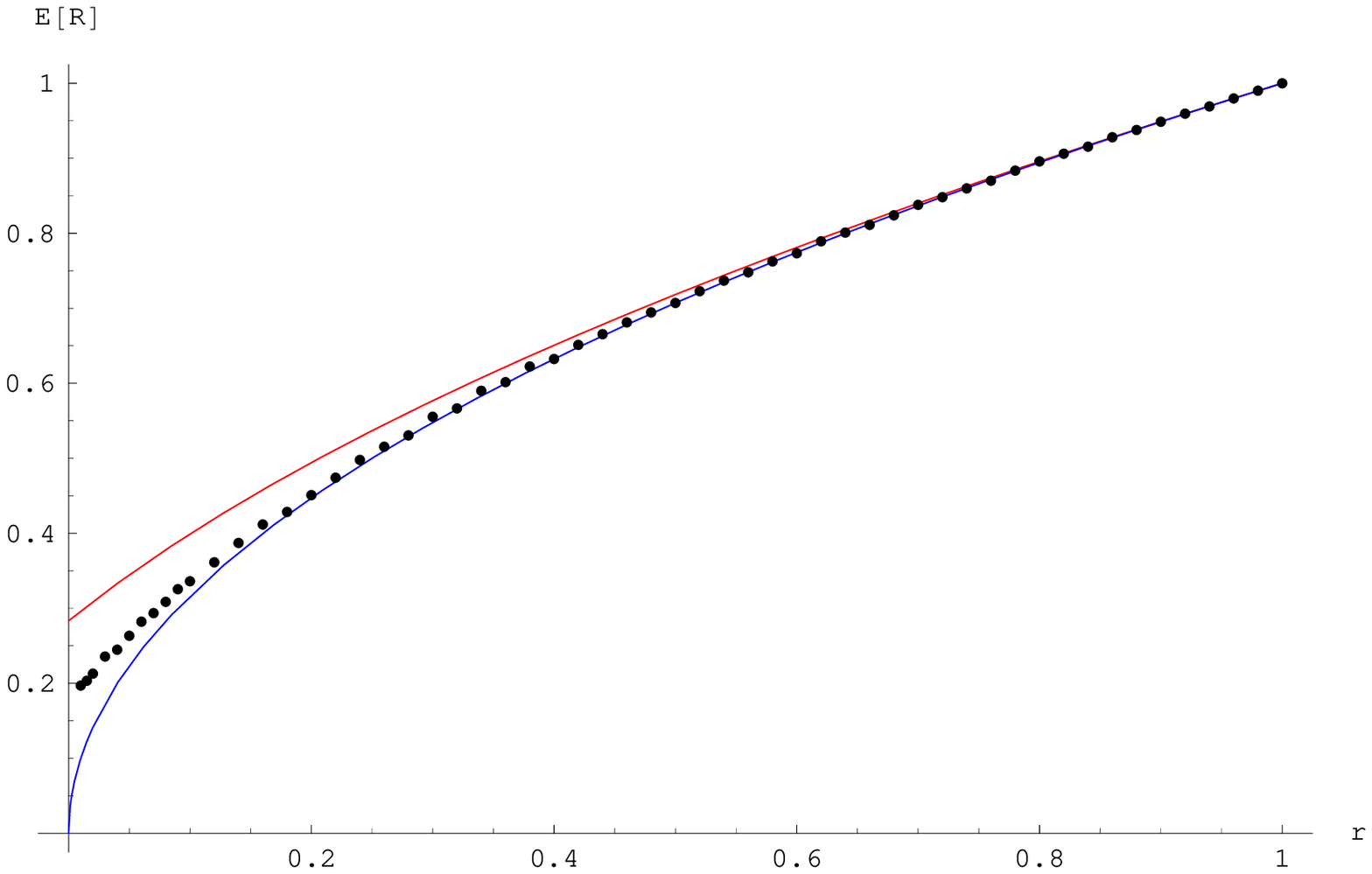}
    }
  }
  \caption{}
  \label{fig:2x2}
\end{figure}

\begin{figure}[H]
  {\psfragscanon\footnotesize
    \psfrag{Empirical Probability}{Probability Density}
    \psfrag{R}{$R$}
    \psfrag{N}{$N$}
    \centerline{\includegraphics[width=0.95\textwidth]{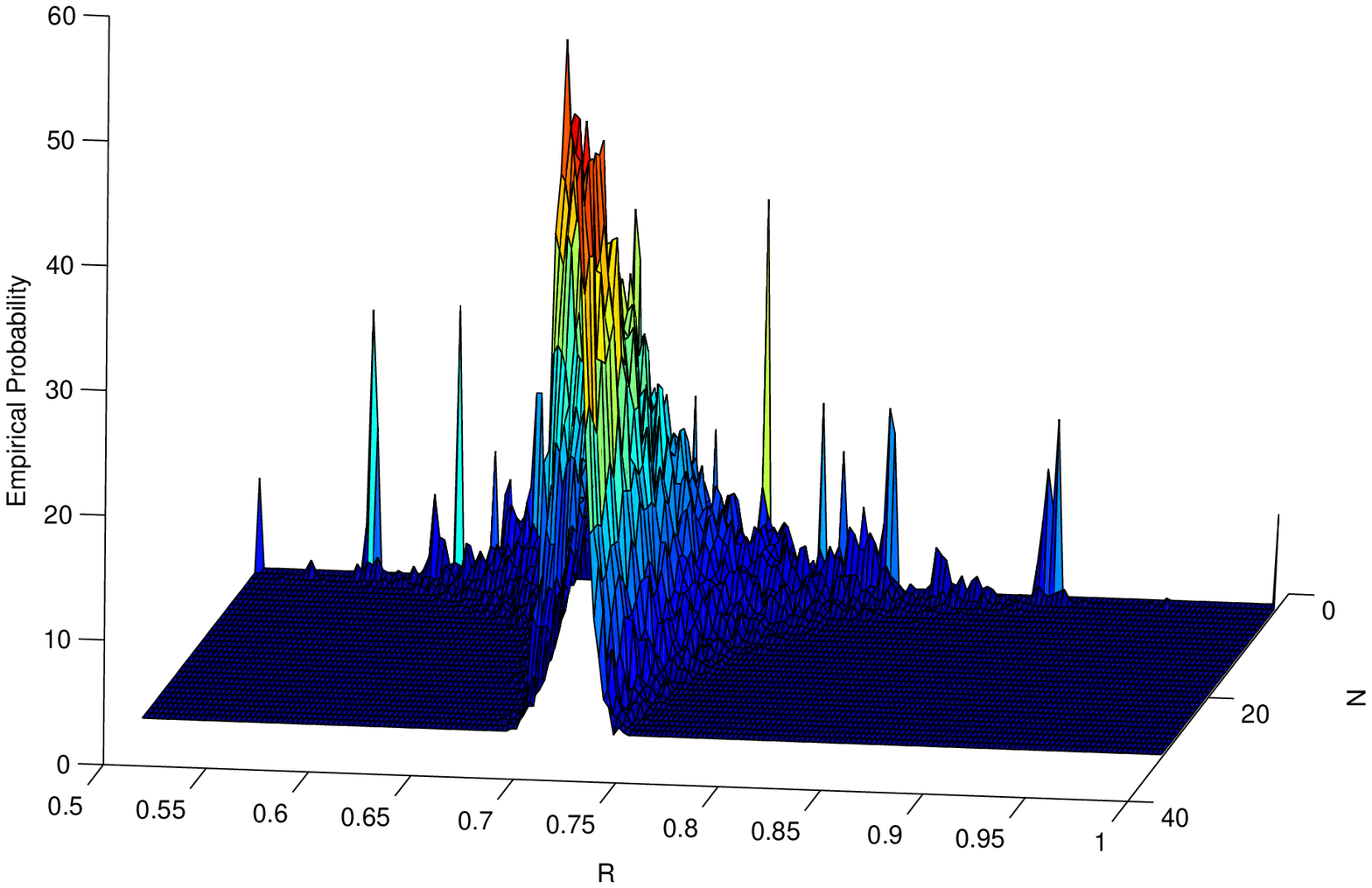}}
  }
  \caption{}
  \label{fig:5}
\end{figure}


\end{document}